\pdfoutput=1
\documentclass[epjST]{svjour}
\usepackage{cite}
\usepackage{tikz}
\usepackage{graphics}
\usepackage{amsmath}

\begin{document}
\title{Non-equilibrium physics of  Rydberg lattices in the presence of noise and dissipative processes}
\author{Wildan Abdussalam \thanks{\email{wildan@pks.mpg.de}} and Laura I. R. Gil}
\institute{Max Planck Institute for the Physics and Complex Systems, N\"othnitzer Stra{\ss}e 38, D-01187 Dresden, Germany 
}
\abstract{We study the non-equilibrium dynamics of driven spin lattices in the presence of decoherence caused by either laser phase noise or strong decay. In the first case, we discriminate between correlated and uncorrelated noise and explore their effect on the mean density of Rydberg states and the full counting statistics (FCS). We find that while the mean density is almost identical in both cases, the FCS differ considerably. The main method employed is the Langevin equation (LE) but for the sake of efficiency in certain regimes, we use a Markovian master equation and Monte Carlo rate equations, respectively. In the second case, we consider dissipative systems with more general power-law interactions. We determine the phase diagram in the steady state and analyse its generation dynamics using Monte Carlo rate equations. In contrast to nearest-neighbour models, there is no transition to long-range-ordered phases for realistic interactions and resonant driving. Yet, for finite laser detunings, we show that Rydberg lattices can undergo a dissipative phase transition to a long-range-ordered antiferromagnetic (AF) phase. We identify the advantages of Monte Carlo rate equations over mean field (MF) predictions.
} %end of abstract

\maketitle
\section{Introduction}
\label{intro}
Non-equilibrium phenomena are ubiquitous in nature and can be found in systems such as fluids~\cite{Cross93}, cells~\cite{berthier13,Prost15}, light harvesting complexes~\cite{Eisfeld11,Valkunas} and polymers~\cite{barbero09}. Similar phenomenology can be studied in controllable artificial systems, in which the presence of driving and decoherence leads to intriguing physics that differs from the equilibrium situation. This has motivated much theoretical \cite{Diehl2011,Mueller2012,Lee2013,Ramos14,Mendoza15} and experimental work \cite{Eisert15,Byrnes14,Blatt12,Rempe91}, based on different experimental platforms ranging from ultracold atoms to driven semiconductor heterostructures. Among such platforms, Rydberg atoms constitute a powerful tool for creating controllable interaction potentials~\cite{ryd_rev,Lahaye2009}. Together with controllable decoherence and coherent driving, this can lead to non-trivial non-equilibrium relaxation \cite{Les2013,Les2014,Marcuzzi14,Guiterrez15,Marcuzzi14-2}, which was also investigated experimentally \cite{schemp14,Malossi2014,Urvoy2015,Valado2015}, and to the new ordered phases \cite{lhc11,hmp13,Qian15,Qian2012,Weimer15,Overbeck16,Maghrebi15,Hoening2014,Lang15,Vermesch15,Chan15,ates12,Hu13} in the limit of strong dephasing. 

Decoherence may arise from laser phase noise or from the strong decay of excited atoms. In the first case, we distinguish between homogeneous or inhomogeneous phase noise: the first one acts globally on the excited states (also referred to as correlated noise), while the second one acts locally on the excited states (uncorrelated noise). Previous works assume that the noise is uncorrelated \cite{Les2013,Les2014,Marcuzzi14,Guiterrez15,Marcuzzi14-2,Robi08,Schonleber14,Honer11,Hernandez08,Kurucz11}.  However, as the laser possesses a spatial correlation length, it can be assumed that laser phase noise leads to homogeneous dephasing. The consequence for the steady states of Rydberg ensembles are yet to be understood.

The presence of strong radiative decays can be a natural means to realise new ordered phases. The interplay between coherent laser excitation and strongly interacting Rydberg atoms can lead to the formation of ordered stationary states \cite{lhc11,hmp13,Qian15,Qian2012,Weimer15,Overbeck16,Maghrebi15,Hoening2014,Lang15,Vermesch15,Chan15,ates12,Hu13}. Previous works have predicted the emergence of steady states with antiferromagnetic order on the basis of mean field theory assuming nearest neighbour (NN) interactions \cite{lhc11}. Yet, large single-site fluctuations related to a simple two-level driving scheme restrict the emergence of such ordering to short length scales in all lattice dimensionalities \cite{hmp13}. Other driving schemes (e.g. in three-level systems) in a 1D setting fail to realise crystallisation \cite{hmp13}, for which long-range order was predicted on the basis of mean field theory in 1D \cite{Qian15} and 2D \cite{Qian2012}. MF predictions are also in conflict with the variational calculations \cite{Weimer15,Overbeck16} and field-theoretical methods \cite{Maghrebi15}, which raises the question whether the emergence of long-range order in dissipative Rydberg lattices is physical or an artifact of the mean-field predictions. This motivates the understanding of a consistent theoretical picture for the possibility of long-range order in dissipative Rydberg lattices.

In this article,  we discuss the dynamics of driven spin lattices in the presence of correlated noise and analyse its difference to the case of uncorrelated noise by comparing the mean density of Rydberg states in the steady state and the full counting statistics (FCS). We show that although the non-equilibrium relaxation for both types of noise is nearly identical, the FCS in the steady differs considerably, which we demonstrate in a few-body scenario. In the presence of strong decay instead of noise, we show that a long-range-ordered AF phase can indeed be realised in dissipative Rydberg lattices when subjected to appropriate coherent driving. In contrast to the equilibrium physics of the corresponding unitary systems, which is well described by mean field models  \cite{wlp08} and NN approximations \cite{jal11}, fluctuations as well as the weak tail of the rapidly decaying interactions are both found to be essential for the physics of the dissipative phase transition.

The article is organised as follows. In Sec. ~\ref{sec:noise}, we describe the dynamics of Rydberg excitations in the presence of noise, where the Rydberg ensembles are modelled as interacting spin-1/2 particles in a one dimensional lattice. For the simple case of two interacting atoms, we distinguish the steady states for correlated and uncorrelated noise. We then extend the system to more atoms and characterise the Rydberg population and the FCS. Next, in Sec. ~\ref{sec:1}, we demonstrate the emergence of long-range order in dissipative Rydberg lattices. We characterise the emergence of AF order and determine the phase diagram in the steady state. We discuss experimental realisations for the described systems in Sec.~\ref{sec:exper}. Finally, Sec.~\ref{sec:conclusion} concludes the article.

\section{Rydberg ensembles in the presence of noise}
\label{sec:noise}
\subsection{1D lattice with two-level driving scheme}
\label{subsec:model}

%%%% figure %%%%%
\begin{figure}[tb]
	\centering
	\includegraphics[width = 100mm]{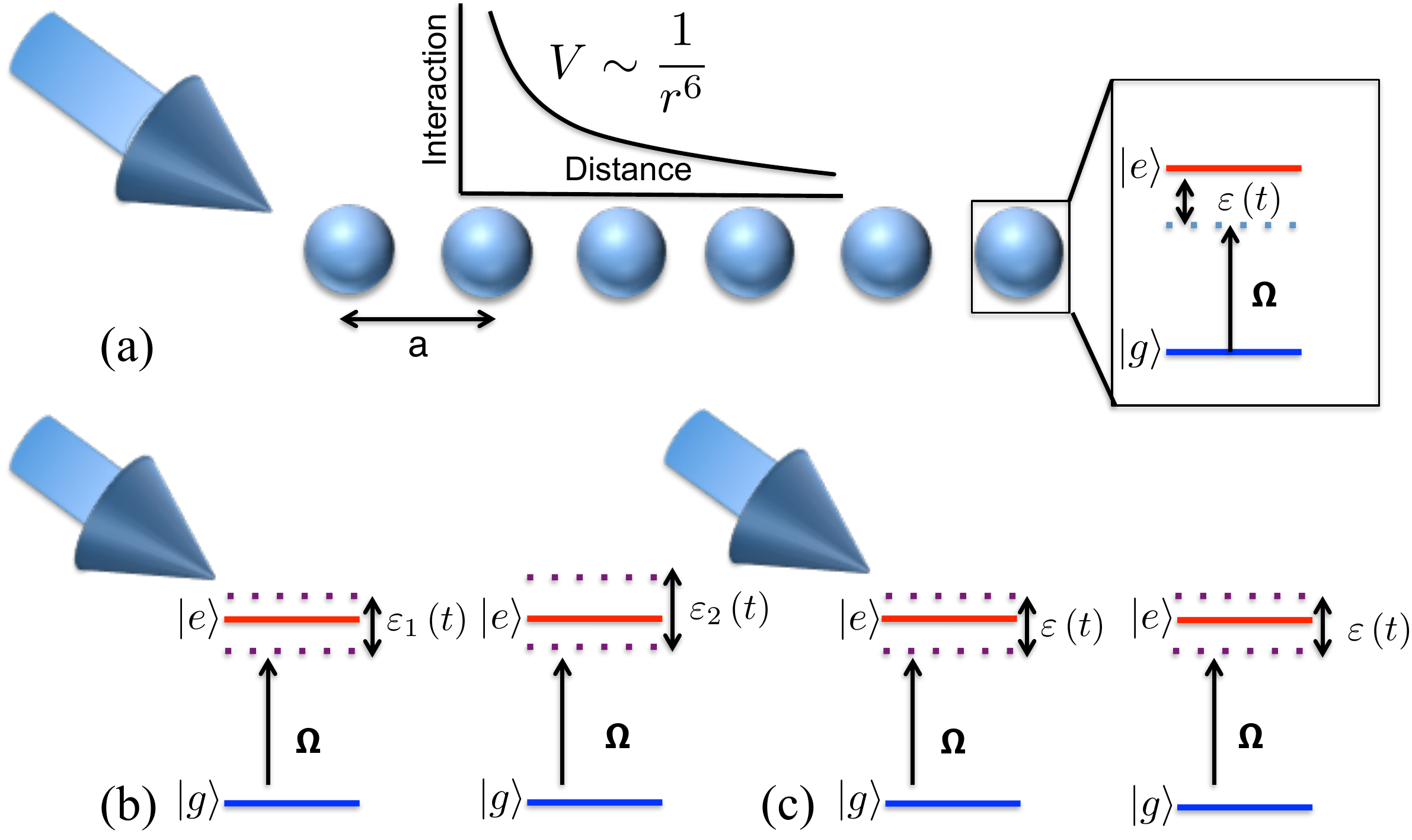}
	\caption{(Color online) (a) Schematics of a one-dimensional lattice in which ground-state atoms $\vert g \rangle$ are laser excited to the Rydberg state $\vert e \rangle$ in the presence of uncorrelated (b) or correlated (c) noise.}
	\label{fig:atomscheme1D}
\end{figure}
%%%% end figure %%%

We consider an ensemble of \textit{N} two-level atoms, each with a ground state $\vert g \rangle$ and a highly excited state $\vert e \rangle$, which are placed in neighbouring sites of a one-dimensional lattice with spacing \textit{a}. The setup is shown in Fig.\ref{fig:atomscheme1D}(a). A pair of excited atoms at sites $i$ and $j$ experiences the repulsive van der Waals (vdW) interaction $V_0/|i-j|^6$ with $V_0=C_6/a^6$ and $C_6 > 0$. Resonant transitions between the two states are driven by a laser with Rabi frequency $\Omega$, which is subjected to phase noise. The general Hamiltonian for this system reads 

%%%% equation %%%%%
\begin{align}
    \label{eq:hamiltionian}
    \hat{H} &= \sum_i \hat{H}_{pn}^{(i)} + \hat{H}_L \\
    \label{eq:localhamiltionian}
    \hat{H}_L &= \frac{\Omega}{2} \sum_i\left(\hat{\sigma}_{eg}^{(i)} + \hat{\sigma}_{ge}^{(i)} \right) +  V_0 \sum_{i < j} \frac{\hat{\sigma}_{ee}^{(i)} \hat{\sigma}_{ee}^{(j)}}{\vert i - j\vert^{6}}
\end{align}
%%%% end equation %%%%% 
where $\hat{\sigma}_{\alpha\beta} = \vert \alpha \rangle\langle \beta \vert$ are transition and projection operators and $\hat{H}_{pn}^{(i)}$ contains the local (l) or global (g) phase noise $\varepsilon(t)$ 
%%%% equation %%%%%
\begin{align}
	\label{eq:locnoise}
	\hat{H}_{pnl}^{(i)} &= \varepsilon^{(i)}\left(t\right)\hat{\sigma}_{ee}^{(i)}\\
	\label{eq:globnoise}
	\hat{H}_{png}^{(i)} &= \varepsilon\left(t\right)\hat{\sigma}_{ee}^{(i)}.
\end{align}
%%%% equation %%%%%
The phase noise $\varepsilon\left(t\right)$ acts as a time-dependent detuning and can be modelled as a one-dimensional Brownian motion \cite{Avan77,Sargent74}. Its time evolution is described by the Langevin equation
%%%% equation %%%%%
\begin{equation}
	\label{eq:noise}
	\dot{\varepsilon} = - \gamma \varepsilon + F(t)
\end{equation} 
%%%% equation %%%%%
where $\gamma$ is a damping term that is inversely proportional to the correlation time $\tau_c = 1/\gamma$. \textit{F(t)} is a random Gaussian function and denotes a rapidly fluctuating force with zero ensemble average $\overline{F(t)} = 0$ and $\overline{F(t)^2}\neq 0$. 
We assume that $F(t)$ has an extremely short correlation time compared to all other characteristic time scales of the system \cite{Avan77,Sargent74}, and thus approximate $\overline{F(t_1)F(t_2)} = 2D \delta \left(t_2 - t_1\right)$. Here \textit{D} describes the magnitude of the fluctuating forces and we assume $\delta\left(t_1-t_2\right)$ as a Lorentzian line shape, which together with the damping $\gamma$ characterises the spectral width $\Gamma$ of Lorentzian line shape \cite{Avan77}
\begin{equation}
	\label{eq:fluctuates}
	 \Gamma = \frac{2D}{\gamma^2} .
\end{equation}
The dynamics of this system is described by the LE in which a single realisation $k$ of the time dependent wave function evolves as $i\vert\dot{\psi}^{(k)}\rangle = \hat{H}\vert\psi^{(k)}\rangle$. For \textit{M} realisations $\vert \psi^{(k)}\rangle$, we calculate the excited state population $\langle\hat{\sigma}_{ee}^{(k, i)}\rangle = \langle \psi^{(k)} \vert \hat{\sigma}_{ee}^{(i)} \vert \psi^{(k)} \rangle$ and take its average
%%%% equation %%%%%
\begin{equation}
	\label{eq:expectation}
	\langle \sigma_{ee} \rangle = \frac{1}{M} \sum_{k=1}^{M}\sum_{i = 1}^{N} \langle \hat{\sigma}_{ee}^{(k, i)}\rangle,
\end{equation}
%%%% equation %%%%%
where $N$ is the atom number. For sufficiently large \textit{M}, the average converges to a constant value, $\langle \sigma_{ee} \rangle$. In order to test the LE, we now compare it to a master equation for the n-body density matrix $\hat{\rho}$ in the Markovian limit. Although the LE is computationally more efficient than the master equation because the corresponding Hilbert space evolves as $2^N$ rather than $2^{2N}$, the master equation is less expensive for the simulation of up to 6 atoms. The time evolution of the density matrix is given by

 \begin{equation}
  \label{eq:markovmastereq}
 \dot{\hat{\rho}} = -i\left[\hat{H}_L, \hat{\rho}\right] + \mathcal{L}\left[\hat{\rho}\right]
 \end{equation}
with $\hat{H}_L$ specified in eq.(\ref{eq:localhamiltionian}). The superoperator $\mathcal{L}$ accounts for the phase noise. In the case of uncorrelated noise it reads
 
\begin{equation}
	\label{eq:supoperloc}
	\mathcal{L}\left[ \hat{\rho} \right]^{(l)} = \Gamma \sum_i \left[\hat{\sigma}_{ee}^{(i)} \rho \hat{\sigma}_{ee}^{(i) } - \frac{1}{2}\lbrace \hat{\sigma}_{ee}^{(i)}\hat{\sigma}_{ee}^{(i)}, \rho\rbrace \right].
\end{equation}
Here the correlation between the phase noise of eq.(\ref{eq:locnoise}) experienced by atoms $i$ and $j$ at two different times  is $\langle \varepsilon^{(i)}\left(t\right) \varepsilon^{(j)}\left(\tau\right)\rangle = (\Gamma/2) \delta_{ij}\delta\left(t-\tau\right)$. For correlated noise [cf. eq.(\ref{eq:globnoise})], the correlation is $\langle \varepsilon\left(t\right) \varepsilon\left(\tau\right)\rangle = (\Gamma/2) \delta\left(t-\tau\right)$ and the superoperator reads

\begin{equation}
	\label{eq:supoperglob}
	\mathcal{L}\left[ \hat{\rho} \right]^{(g)} =  \Gamma \left[\hat{\Sigma}_i \hat{\rho} \hat{\Sigma}_j - \frac{1}{2}\lbrace \hat{\Sigma}_i\hat{\Sigma}_j, \hat{\rho}\rbrace \right]
\end{equation}
where $\hat{\Sigma}_i = \sum_{i}\hat{\sigma}_{ee}^{(i)}$. The mean density $\rho\left(\tau\right)$ is obtained by taking the trace $ Tr\lbrace \hat{\rho} \hat{\Sigma}_i\rbrace / N$ which is equivalent to $\langle \sigma_{ee} \rangle/N$ in eq. (\ref{eq:expectation}).  

%%%%%%%%%%%%%%%%%%%%%%%%%%%%%%%%%%%%%%%%%
\begin{figure}[tb]
   \centering
   \includegraphics[width = 100 mm]{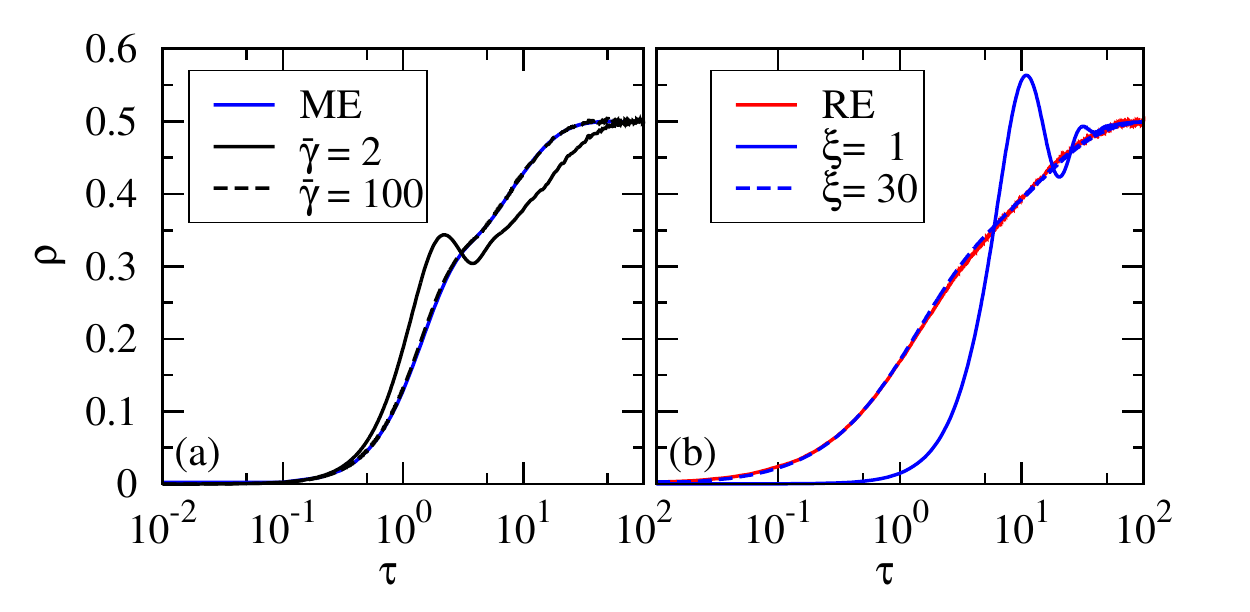} 
   \caption{(Color online) The averaged mean density as a function of time (a) in the presence of correlated noise for $N = 2$, $\xi = 4$, $R = 1$ and $M = 10^4$ calculated by the LE for two different $\bar{\gamma}$ compared to the Markovian master equation (ME). (b) The averaged mean density for $N = 6$ and $R = 1$ calculated by the classical rate equation (RE) with $M = 10^4$ compared to the Markovian master equation with two different $\xi$ in the presence of uncorrelated noise.}
   \label{fig:fitmethod}
\end{figure}
%%%%%%%%%%%%%%%%%%%%%%%%%%%%%%%%%%%%%%%%%%

We introduce the dimensionless rescaled time $\tau = (4\Omega^2\big{/}\Gamma) \times t$, as in \cite{Les2013}. The many-body state is determined by three independent dimensionless parameters:  the spectral width of the noise $\xi = \Gamma\big{/}\Omega$, the damping term $\bar{\gamma} = \gamma\Gamma\big{/}4\Omega^2$, and the interaction strength parameter $R^{6} = \left[ C_{6} \big{/} \left(\Gamma a\right) \right] $. 

Fig. \ref{fig:fitmethod} (a) shows the relaxation of the mean density in the presence of correlated noise. We show that in the limit $\bar{\gamma} >> 1$, i.e. for very short correlation time, the LE is in excellent agreement with eq. (\ref{eq:markovmastereq}). Later on in subsec.~\ref{subsec:Nbody}, we will discuss the dynamics in the strong coupling limit. There, the classical rate equation is computationally more efficient than the LE. For sufficiently strong dephasing, the quantum dynamics can be reduced to the diagonal elements of $\rho$ upon adiabatic elimination of its coherences \cite{Ates2007}. This simplifies the time evolution to a classical rate equation model for the joint probabilities $\rho_{S_1, ...., S_N}$ of an atom \textit{i} in the Rydberg state $(S_i = 1)$ or in the ground state $(S_i = 0)$. The time evolution is determined by

\begin{eqnarray}
\label{eq:rate_equ}
&&\dot{\rho}_{S_1,...,S_N}\!\!=\!\!\sum_i \Bigl[(1-S_i)D(\delta_i)+S_i P(\delta_i)\Bigr]\rho_{S_1,...,1-S_i,...,S_N}\nonumber\\
&&\quad -\Bigl[(1-S_i)P(\delta_i)+S_i D(\delta_i)\Bigr]\rho_{S_1,...,S_i,...,S_N}\;,
\end{eqnarray}
where $P\left(\delta_i\right)$ and $D\left(\delta_i\right)$ denote the excitation and de-excitation rates, respectively. The rates can be expressed as $ P\left(\delta_i\right) = 1 \big/ \left[4\left(1 + \delta_i^2\right)\right]$  and $D\left(\delta_i\right) = 1 \big/ \left[4\left(1 + \delta_i^2\right)\right]$. The interaction enters through an effective frequency detuning $\delta_i=\delta-R^{6}\sum_{j\neq i}S_j|{\bf r}_i-{\bf r}_j|^{-6}$ which accounts for the level shift of the \textit{i}th atom due to its surrounding Rydberg excitations. 

We use the Markovian master equation as a test for the classical rate equation. As shown on Fig.\ref{fig:fitmethod}(b), for $\xi \gg 1$ the mean density is indeed well reproduced by the Markovian master equation model (eq.~\ref{eq:markovmastereq}). This is in agreement with the results in \cite{Ates2007,Olmos14,Mattioli15}. In the following, we will only consider short correlation times $\bar{\gamma}\gg1$ and strong dephasing $\xi \gg 1$. 

\subsection{Two interacting atoms in the presence of correlated and uncorrelated noise}
\label{subsec:twoatoms}

In this subsection, we investigate the difference between correlated and uncorrelated noise for two atoms, with emphasis on the steady-state distribution in the non-interacting and interacting case. 

\begin{figure}[tb]
   \centering
   \includegraphics[width = 80 mm]{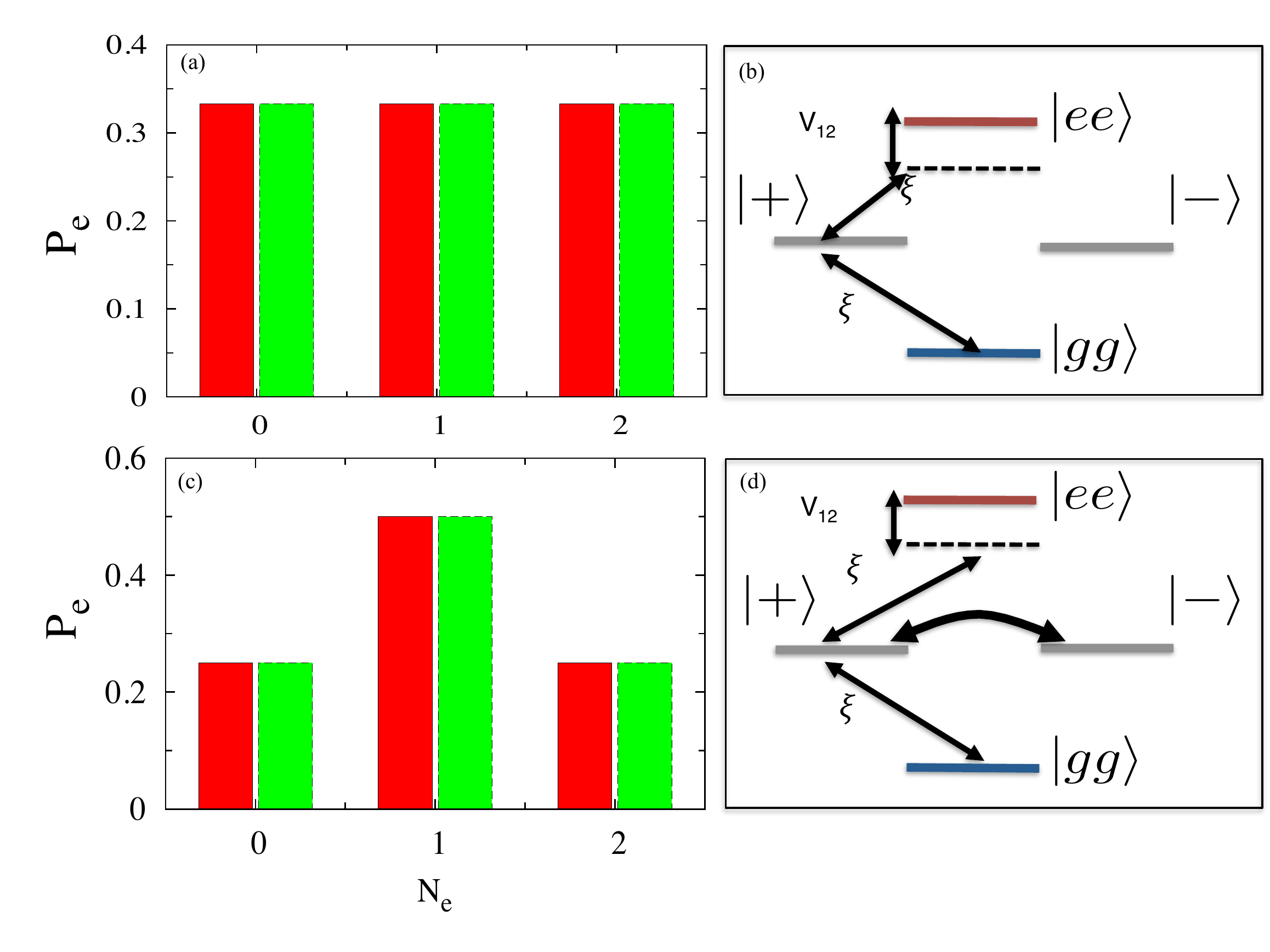} 
   \caption{(Color online) The excitation probability $P_e$ in the steady state as a function of the number of excitations $N_e$ in the presence of correlated (a) and uncorrelated (c) noise. Red and green bars indicate non-interacting and interacting Rydberg atoms, respectively. (b) and (d) illustrate the couplings between the two-atom energy states in the presence of correlated and uncorrelated noise.}
   \label{fig:two-atom-elements}
\end{figure}

For non-interacting atoms in the presence of correlated noise [see Fig.~\ref{fig:two-atom-elements} (a) (red bar)], the distribution is uniform for any excitation number. This can be understood from [Fig.~\ref{fig:two-atom-elements} (b) assuming no interaction], the phase $\varepsilon\left(t\right)$ acquired by each atom is identical. Consequently, the coherent driving couples the ground state of atoms to the symmetric state $\vert + \rangle = \left(1/\sqrt{2}\right)\left(\vert ge \rangle + \vert eg \rangle\right)$ and the latter to the doubly excited state. However, the antisymmetric state  $\vert - \rangle =  \left(1\big / \sqrt{2}\right)\left(\vert ge\rangle - \vert eg \rangle\right)$ is decoupled. Thus, the excitation probability $P_e$ for any excitation number $N_e$ is $P_e=1/3$. 

On the contrary, the steady-state distribution in the presence of uncorrelated noise exhibits a non-uniform distribution [see Fig.~\ref{fig:two-atom-elements} (c) (red bar)]. This can be understood from Fig.~\ref{fig:two-atom-elements} (d): each atom acquires a different phase $\varepsilon^{(i)}\left(t\right)$. Consequently, the symmetric state $\vert + \rangle$ couples to the antisymmetric state $\vert - \rangle$ which leads to a different population distribution with $P_0 = 1{\big /}4$, $P_1 = 1{\big /}2$, and $P_2 = 1{\big /}4$.

%%%%%%%%%%%%%%%%%%%%%%%%%%%%%%%%%%%%%%%%%
%%%%%%%%%%%%%%%%%%%%%%%%%%%%%%%%%%%%%%%%%%

In the presence of interactions, the steady-state distribution is the same as in the non-interacting case [see Fig. \ref{fig:two-atom-elements} (a) and (c) green bar]. Despite the presence of Rydberg-Rydberg interactions, the steady state for both types of noise remains unchanged, as the interaction only shifts the energy of the doubly-excited state. In the presence of correlated noise, the energetic shift of the doubly-excited state is unable to break the symmetry, resulting in the same distribution as in the non-interacting case. For the minimal example of two interacting atoms, we have shown that the steady-state distribution changes remarkably with the type of noise considered, yet it is unchanged in the case of non-interacting and interacting atoms. In the next subsection we investigate whether this finding persists in larger ensembles.  

%%%%%%%%%%%%%%%%% New sub section %%%%%%%%%%%%%%%%%
\subsection{Few-body simulations in the presence of correlated and uncorrelated noise}
\label{subsec:Nbody}
We extend the system size to 6 atoms. We determine the mean density  $\rho\left(\tau\right)$ in the presence of correlated and uncorrelated noise for various ranges of the interaction strength. We classify the interaction strength $R^6$ into weak interactions for $\left(R^6 \ll \xi \right)$, intermediate interactions for $\left(R^6 \leq \xi\right)$ and strong interactions for $\left(R^6 > \xi\right)$. Increasing the interaction strength between the atoms slows down the relaxation time. Therefore, we will use three different methods for the three different limits : the Markovian master equation for the weak and intermediate interactions in the Markovian limit ($\bar{\gamma}\gg 1$), the LE for intermediate and strong interactions and arbitrary values of $\bar{\gamma}$ and $\xi$, and the classical rate equation for strong interactions and dephasing ($\xi \gg 1$) in the Markovian limit. 

%%%%%%%%%%%%%%% Figure %%%%%%%%%%%%%%%
\begin{figure}[tb]
   \centering
   \includegraphics[width = 100mm]{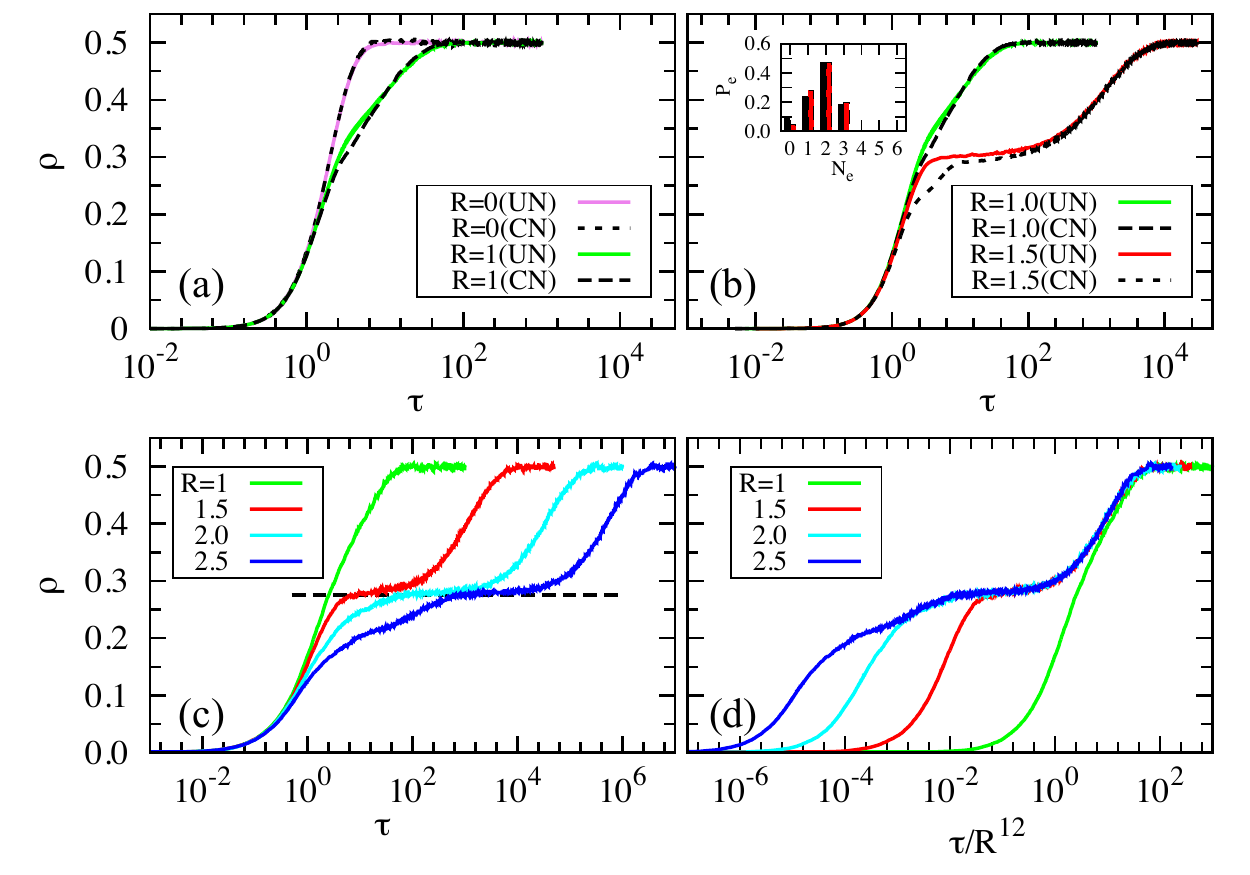} % requires the graphicx package
   \caption{(Color online) Relaxation of the mean density of Rydberg atoms in a one-dimensional system for \textit{N} = 6 and $M=10^4$ calculated by the LE [(a), (b)] and the rate equation [(c), (d)]. The comparison between correlated (CN) and uncorrelated (UN) noise with $\xi=4$ and $\bar{\gamma}=100$ for non-interacting and interacting atoms is shown in (a) and for intermediate as well as strong interactions in (b). The inset in (b) shows the excitation distribution at the plateau for $R = 1.5$. Relaxation in the presence of uncorrelated noise for different strong interactions is presented in (c), where the dashed line indicates the mean density of hard dimers. For long times the steady state is approached exponentially at a rate $\propto R^{-12}$ (d).}
   \label{fig:dyn6atomsMC}
\end{figure}
%%%%%%%%%%%%%%%%%%%%%%%%%%%%%%%%%%%

Fig. \ref{fig:dyn6atomsMC} shows the relaxation of the mean density for $N = 6$ in the limit of intermediate and strong interactions. The initial state is $\rho\left(0\right) = 0$ which corresponds to all atoms in the ground state. In the non-interacting case [see Fig. \ref{fig:dyn6atomsMC} (a)], the mean density shows identical relaxation behaviour for both types of noise that reaches the steady state, which is a mixed state $\rho\left(\infty\right)\simeq0.5$. As $R^6$ is increased to the intermediate interaction limit, it slows down the relaxation because an excited atom suppresses other excitations in neighbouring sites.

We now consider an interaction strength of $R^6 > \xi $. In Fig.~\ref{fig:dyn6atomsMC} (b), we show that although the interaction strength is three times larger than the spectral width of the laser ($R^6 > \xi$), the mean density relaxation is nearly identical for both types of noise. A small discrepancy appears between correlated and uncorrelated noise for intermediate interaction before entering the plateau which we will discuss in the next subsection. The plateau emerges at $\rho_{plat} \approx 0.27$  just before the system relaxes exponentially to the steady state. In the range of time at which a plateau is present the excitation distribution for both types of noise is dominated by two excited atoms [see Fig.~\ref{fig:dyn6atomsMC}(b) inset]. This shows the strong suppression of simultaneous excitation of neighbouring atoms. The plateau density corresponds to hard dimers whose a value $\rho_{dim} = (1-1/\sqrt{5})/2 \sim 0.276$ \cite{Les2011} and is in agreement with \cite{Les2013} where only uncorrelated noise was considered. As we increase the interaction strength up to $R = 2.5$ [see Fig. \ref{fig:dyn6atomsMC} (c) and (d)], the plateau value of $\rho_{dim} \sim 0.276$ is confirmed. This shows that the typical relaxation behaviour in the presence of phase noise, which was discussed for very large ensembles in \cite{Les2013}, can instead be implemented with a small number of atoms \cite{Labuhn15,Beguin2013}.

Although the dynamics of the mean density for both types of noise is nearly identical, the relaxation to the transient state shows a small discrepancy that due to different excitation distributions. Therefore, in the next subsection, we will investigate the atom counting statistics for both types of noise.

\subsection{Atom counting statistics in the presence of correlated and uncorrelated noise}
\label{subsec:qmandel}
We now discuss the atom counting statistics for a small number of atoms with $N \geq 6$. One can quantify the distribution in terms of the {\bf variance} $\left(\Delta  \rho \right)^2$ which corresponds to the fluctuations of a statistical distribution about its mean value. The variance is equal to the square of the standard deviation $ \Delta \rho $ and is defined by:

\begin{equation}
	\label{eq:variance}
	\left(\Delta \rho\right)^2 = \sum\limits_{i}^{N} \frac{\langle \sigma_{ee}^{(i)2}\rangle - \langle \sigma_{ee}^{(i)}\rangle^2}{\langle \sigma_{ee}^{(i)}\rangle}.
\end{equation}
Depending on the relation between the variance and the mean value one can distinguish three cases. In a sub-Poissonian distribution, the variance is less than the mean value $ \left(\Delta \rho\right)^2  <  \rho $. In a Poissonian distribution, $ \left(\Delta \rho\right)^2 =  \rho $. In a super-Poissonian distribution, $\left(\Delta \rho\right)^2  >   \rho $~\cite{Foxbook}. 

%%%%%%%%%%%%%%%%%%%%%%%%%%%%%%%%%%%%%%%%%
\begin{figure}[tb]
   \centering
   \includegraphics[width = 100mm]{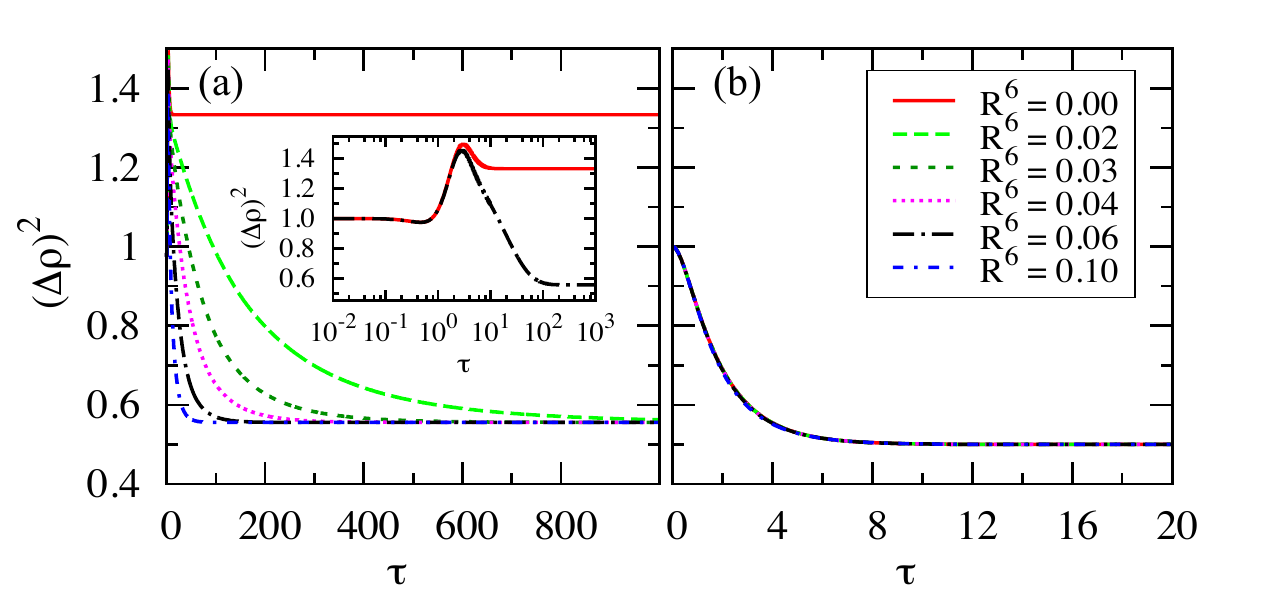} % requires the graphicx package
   \caption{(Color online) Variance $\left(\Delta \rho\right)^2$ as a function of time for $N=6$, $\xi = 5$ and various $R^6$ in the presence of correlated noise (a) and uncorrelated noise (b) calculated by the ME. Inset shows the relaxation in logarithmic x-axis scale for $R^6=0$ and $R^6=0.06$.}
   \label{fig:slopeQ}
\end{figure}
%%%%%%%%%%%%%%%%%%%%%%%%%%%%%%%%%%%%%%%the steady-state variance%%

In order to reach the steady state within a short relaxation time, we have performed the simulation in the weak interaction limit. In Fig.~\ref{fig:slopeQ} we show the variance for various interaction strengths $R^6$ in the presence of correlated (a) and uncorrelated (b) noise. For correlated noise [see Fig.~\ref{fig:slopeQ} (a) inset], the relaxation behaviour and steady-state variance for $R=0$ are fundamentally different from an interacting system with $R \neq 0$. This is due to the fact that for $R=0$ the ground state only couples to the symmetric states, i.e. Dicke states~\cite{Dicke54}, yet is decoupled from the non-symmetric states. For $R \neq 0$, the system is no longer decoupled from the non-symmetric states, instead the symmetric states with $N_e \geq 2$ are shifted due to Rydberg-Rydberg interactions, resulting in a symmetry breaking and depopulation of symmetric states into non-symmetric states. This leads to a different steady-state variance compared to $R=0$. Furthermore, a slight increase of the weak interaction strength leads to strongly reduced relaxation times as long as $R^6 \ll \xi$ [see Fig.~\ref{fig:slopeQ} (a)]. In contrary, the variance in the presence of uncorrelated noise shows a completely different relaxation behaviour and reaches a different steady state. As shown in Fig.~\ref{fig:slopeQ} (b), the variance relaxation and the steady state are completely independent of the interaction strength for the small energetic shifts considered. This is due to the fact that for uncorrelated noise in the absence of interactions, each atom experiences a different time-dependent detuning, resulting in a coupling between symmetric and non-symmetric states. In this weak interaction limit, the relaxation time for correlated noise is longer than for uncorrelated noise. 

\begin{figure}[tb]
   \centering
   \includegraphics[width = 100mm]{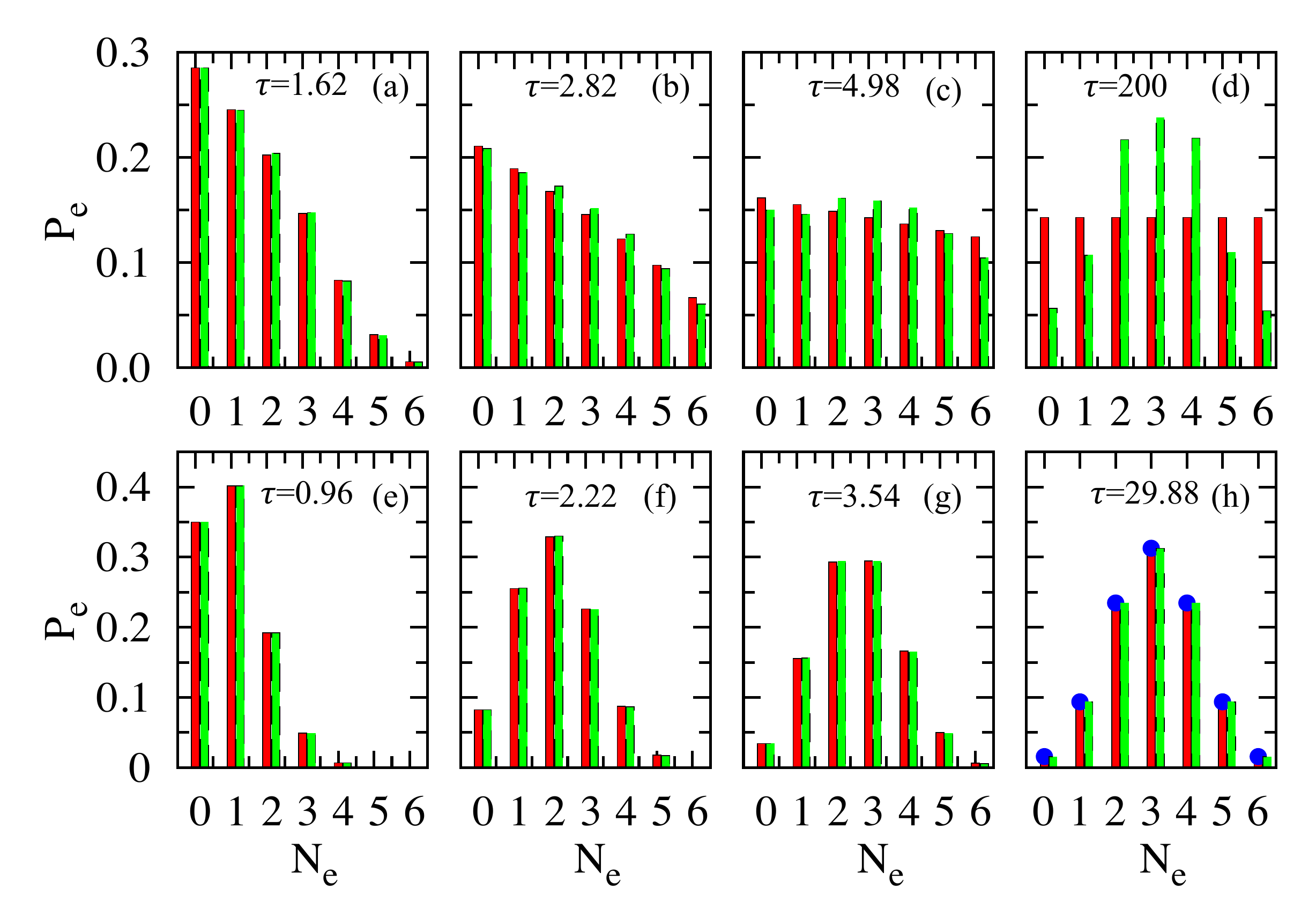} % requires the graphicx package
   \caption{(Color online) Dynamics of the excitation probability $P_e$ as a function of the excitation number $N_e$ for the non-interacting (red bars) and interacting (green bars) case in the presence of correlated (a-d) and uncorrelated (e-h) noise. Panel (a-d) corresponds to the parameters in Fig.~\ref{fig:slopeQ} (a)-inset and (e-h) corresponds to the parameters in Fig.~\ref{fig:slopeQ}(b) for $R^6=0$ and $R^6=0.06$.}
   \label{fig:distribution_dyn}
\end{figure}

We now discuss the dynamical evolution of the excitation distribution and classify the type of distribution in the steady state according to the criteria mentioned above. We show the probabilities $P_e$ for finding $N_e$ excitations at different times in Fig.~\ref{fig:distribution_dyn}, comparing the correlated noise with uncorrelated noise. For correlated noise in the absence of interactions, as shown in Fig.~\ref{fig:distribution_dyn} (a-d), the ground state population $\vert gg ... g\rangle$ slowly migrates to higher excitation numbers only via the symmetric states, resulting in uniform distribution in the steady state. There, $P_e\simeq 1{\big /}\left(N+1\right)$ for any excitation number. For example, for $N = 6$ this results in $P_e\simeq 1{\big /}7$ for any number of excitations. Thus, the variance can be calculated analytically for any arbitrary number of atoms $N$. For example, for $N = 6$ the variance is $\left(\Delta \rho\right)^2 = 4/3$ [see Fig.~\ref{fig:slopeQ} (a)]. In the presence of interactions, the migration via the symmetric states is followed by the migration into non-symmetric states, giving rise to a non-uniform distribution in the steady state. As shown in Fig.\ref{fig:distribution_dyn}(d), the non-uniform distribution corresponds to a super-Possionian distribution in which the steady-state variance $\left(\Delta \rho\right)^2 > \rho =0.5$. 

%%%%%%%%%%%%%%%%%%%%%%%%%%%%%%%%%%%%%%%%%
%%%%%%%%%%%%%%%%%%%%%%%%%%%%%%%%%%%%%%%the steady-state variance%%

Fig. \ref{fig:distribution_dyn} (e-h) shows the excitation distribution in the presence of uncorrelated noise. The migration shows different behaviour from correlated noise. In the absence of interactions, since the symmetric state is already coupled to non-symmetric states, the ground state population rapidly migrates to higher excitation numbers, shown by the rapidly reduced population of the ground state. When weak interactions are present, the relaxation is unchanged. The steady-state variance exhibits a Poissonian distribution for both cases ($\left(\Delta \rho\right)^2  =  \rho=0.5$), shown in Fig.~\ref{fig:distribution_dyn} (h) [red and green bars]. The Poisson distribution can be expressed as

\begin{equation}
\label{eq:binomialdist}
f\left(N_e, N, \rho\left(\infty\right)\right) = \binom{N}{N_e} \rho(\infty)^{N_e} \left(1 - \rho\left(\infty\right)\right)^{N-N_e}. 
\end{equation}
For example, for $N_e = 3$ and $N =6$ one gets $f\left(N_e, N, \rho\left(\infty\right)\right) \simeq 0.3125$, in agreement with a simulation result [Fig.~\ref{fig:distribution_dyn} (h-red and green bars)]. 
%%%%%%%%%%%%%%%%%% steady state variance %%%%%%%%%%%%%%%%%%%%%%%
\begin{figure}[tb]
   \centering
   \includegraphics[width = 90mm]{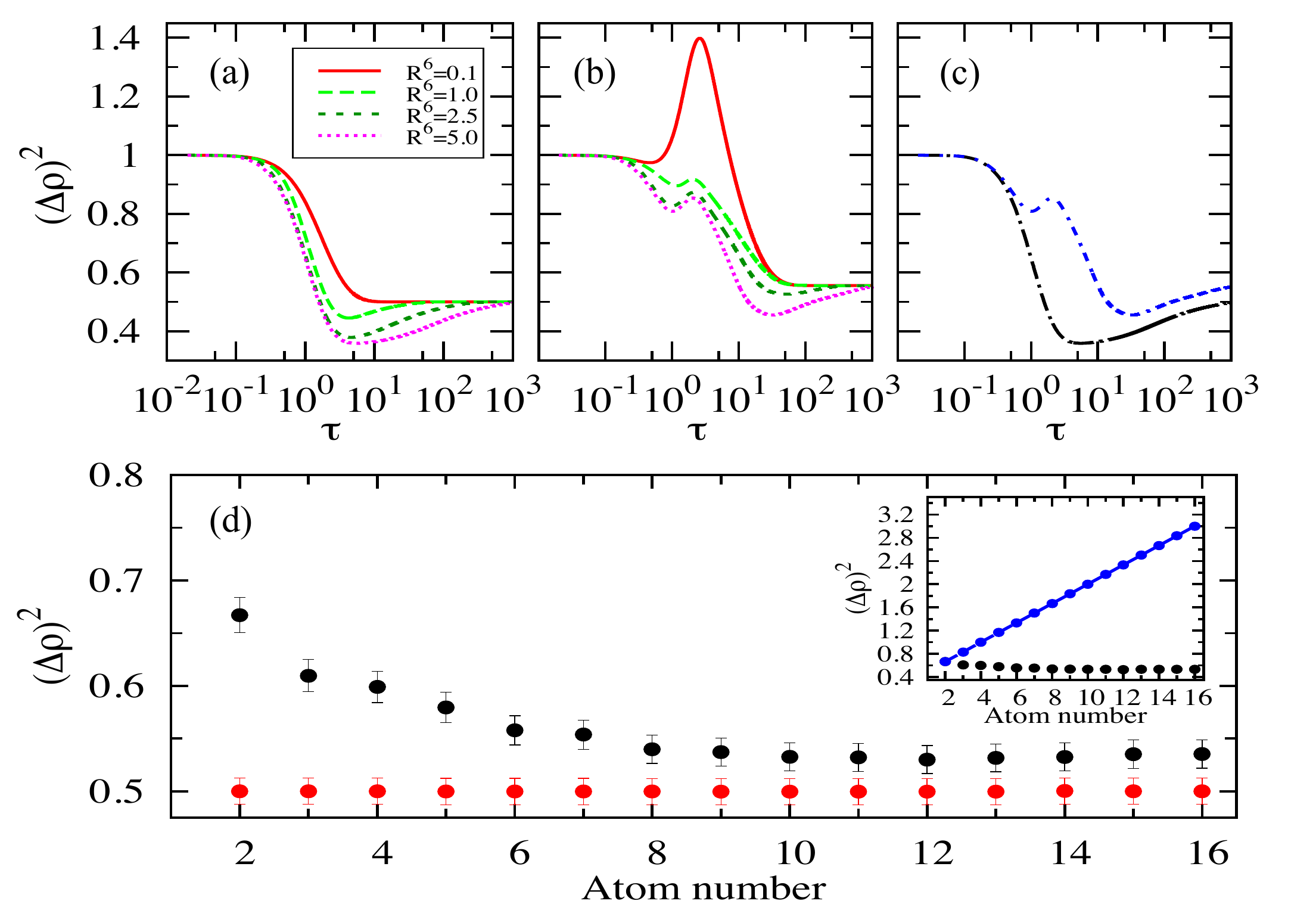} 
   \caption{(Color online) Variance as a function of time for $\xi = 5$ in the limit of intermediate interactions and the presence of (a) uncorrelated noise or (b) correlated noise. (c) shows a comparison of both cases [correlated noise (blue dashed) and uncorrelated noise (black dashed)] for $R^6=5$ calculated by the ME. Panel (d) shows the steady-state variance as a function of the atom number $N$ for interacting atoms in the presence of correlated (black dots) and uncorrelated (red dots) noise calculated by the LE for $5\times10^4$ realisations. In the presence of correlated noise (inset), non-interacting atoms (blue dot-dashed) show a linear increase of the variance with the atom number (exponent : $0.167 \pm 0.01$) while for interacting atoms the variance decreases.}
   \label{fig:slopeQsteady}
\end{figure}
%%%%%%%%%%%%%%%%%%%%%%%%%%%%%%%%%%%%%%%%%%%%%%%%%%%%%%%
In the limit of intermediate interaction strengths, the coupling for uncorrelated noise starts to show a dependence on the interaction strength. As shown in Fig.~\ref{fig:slopeQsteady}(a), the increase of interaction strength $R^6$ slows down the variance relaxation. The slow down is followed by the emergence of a dip that corresponds to the transition of the mean density to the transient state in Fig.~\ref{fig:dyn6atomsMC}. For correlated noise [see Fig.~\ref{fig:slopeQsteady} (b)], as we enter the intermediate interaction regime $\left(R^6 \leq \xi\right)$, the interaction strength starts to slow down the relaxation and the dip emerges at a later time than for uncorrelated noise [see Fig.~\ref{fig:slopeQsteady} (c)]. This means that correlated noise slows down the transition to the transient state compared to uncorrelated noise. 

In the presence of correlated noise, we have seen that the steady-state variance in a chain of 6 atoms differs for non-interacting and interacting cases.  An interesting question is whether this difference depends on the atom number. Fig. \ref{fig:slopeQsteady}(d)[inset] shows the comparison of variances between non-interacting and interacting atoms. For the non-interacting case, it is interesting to note that the steady-state variance increases linearly with $N$. As the interaction is switched on, the fluctuating excitations are strongly suppressed shown by the decrease of variance with increasing number of atoms that converges to a value $\left(\Delta\rho(\infty)\right)^2 > 0.5$, slightly above the steady-state in the presence uncorrelated noise [see Fig.~\ref{fig:slopeQsteady}(d)]. The different steady states are due to the fact that for correlated noise the symmetric state for $N_e = 1$ only gets shifted by the global noise. For uncorrelated noise, the distribution follows a Poissonian distribution for the non-interacting and interacting case, resulting in a constant variance for any number of atoms.

In this section, we have investigated the difference between correlated and uncorrelated noise deriving from laser phase noise. In the next section [sec.~\ref{sec:1}], we study the non-equilibrium physics in the presence of a different dissipative process in a two-dimensional lattice that originates from the spontaneous decay of atoms. We show that this can be a natural means to realise ordered phases.
%%%%%%%%%%%%%%%%%%%%%%%%%%%%%%%%%%%%%%%%%%%%%%%%%%%%%%%
									
%%%%%%%%%%%%%%%%%%%%%%%%%%%%%%%%%%%%%%%%%%%%%%%%%%%%%%%
\section{Antiferromagnetic long-range order in dissipative Rydberg lattices}
\label{sec:1}
In the previous section, we have considered the dynamics and steady state of a one-dimensional lattice of two-level atoms in the presence of laser phase noise. There, we have found intriguing relaxation behaviour and steady-state distributions. An interesting question is: How do the dynamics and steady state behave when the decoherence originates from spontaneous decay instead of laser phase noise? In particular, we are interested in the conditions to realise AF long-range order. Previous work has investigated AF order of the steady state in a 1D setting assuming nearest-neighbour (NN) interactions \cite{hmp13}. However, the large single-site fluctuations associated with a simple two-level driving scheme restricts the emergence of that ordering to short length scales for all spatial lattice dimensions. Moreover, simulations in 1D showed that long-range crystallisation is also prevented for other driving schemes. Therefore, we investigate the dynamics of the system by means of a three-level driving scheme in a square lattice. Part of the results presented in this section have been published in~\cite{Hoening2014}.

\subsection{2D lattice with three-level driving scheme}
\label{subsec:afmodel}
\begin{figure}[tb]
   \centering 
   \includegraphics[width = 100 mm]{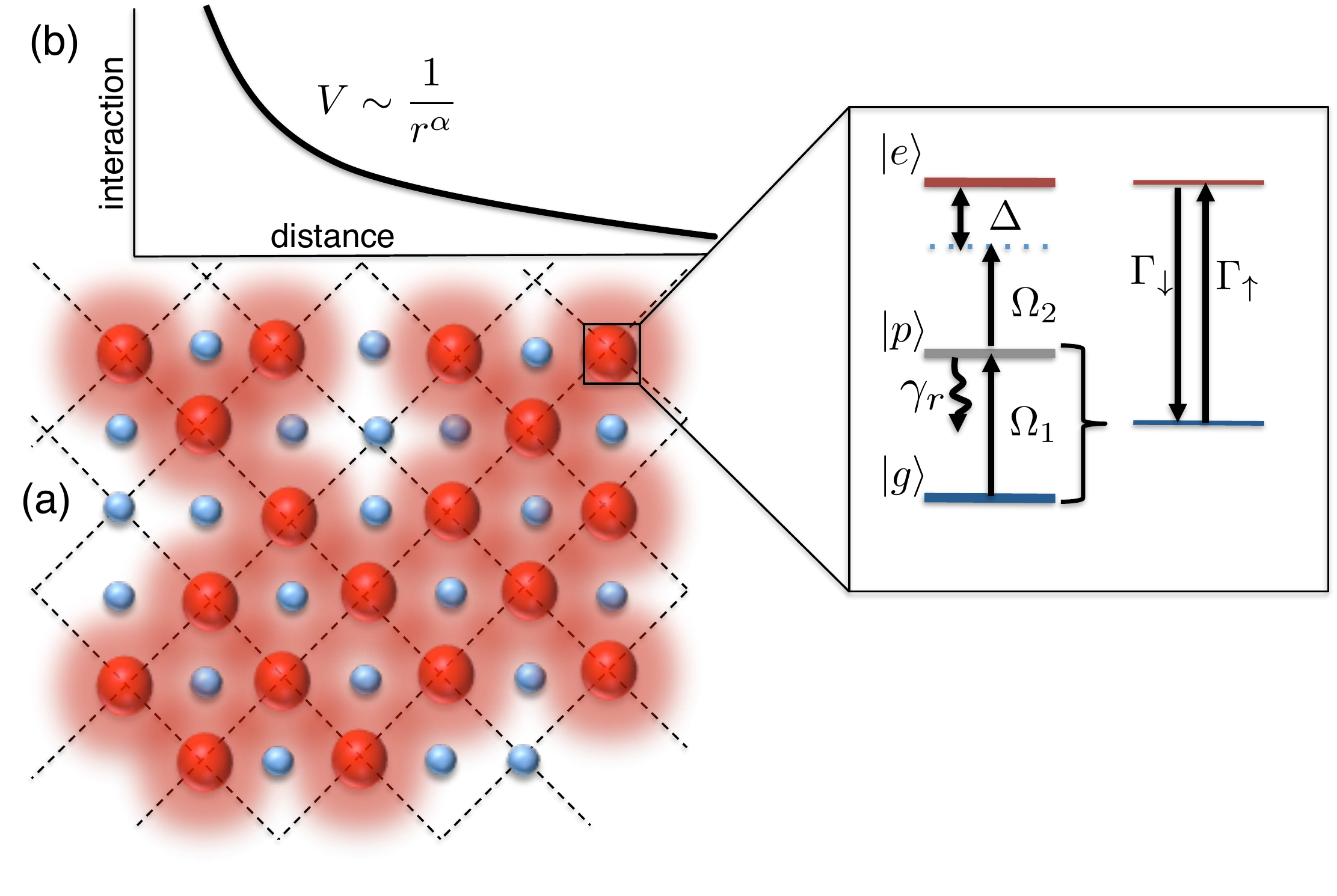} 
   \caption{(Color line) (a) Schematics of a two-dimensional lattice in which ground-state atoms (small blue spheres) are laser excited to Rydberg states (large red spheres). The interplay of dissipation and Rydberg-Rydberg interactions (b) can give rise to antiferromagnetic long-range order, where excitations predominantly occupy one checkerboard sublattice. A possible realisation of such effective two-level systems with tunable excitation rates $\Gamma_{\uparrow}$ and  $\Gamma_{\downarrow}$ is illustrated in (c)(see text for details). Figure adapted from~\cite{Hoening2014}.} 
   \label{fig1}
\end{figure}

We consider laser-driven atoms on a quadratic lattice of length \textit{L}. An atom is modelled by a three-level ladder-type system which is excited and de-excited by two lasers with different Rabi frequencies $\Omega_1$, $\Omega_2$ and detuning $\Delta$. The excited (Rydberg),  intermediate, and  ground states of atom \textit{i} are denoted by $\vert e^{(i)} \rangle$, $\vert p^{(i)} \rangle$ and $\vert g^{(i)} \rangle$, respectively [see Fig.\ref{fig1} (a)]. The Hamiltonian which governs the system reads

%%%%%%%%%%%%%%%%%%%%%%%%%%%%%%%%%%%%%%%%%%%%%%%%%%%%%%
%%%%%%%%%%%%%%%%%%%%%%%%%%%%%%%%%%%%%%%%%%%%%%%%%%%%%%

\begin{align}
	\label{eq:ham}
	\hat{H}&=\sum_{i}\hat{H}_{i} + V_0\sum_{i<j}\frac{\hat{\sigma}_{\rm ee}^{(i)}\hat{\sigma}_{\rm ee}^{(j)}}{|{\textbf r}_i-{\textbf r}_j|^{\alpha}}, \\ 
	\hat{H}_i&= \frac{\Omega_1}{2}\left(\hat{\sigma}_{pg}^{(i) }+ \hat{\sigma}_{gp}^{(i)}\right) + \frac{\Omega_2}{2}\left(\hat{\sigma}_{ep}^{(i) }+ \hat{\sigma}_{pe}^{(i)}\right) -\Delta\hat{\sigma}_{\rm ee}^{({i})}
	\label{eq. locahamAF} 	
\end{align}
where $\hat{\sigma}_{\alpha\beta} = \vert\alpha\rangle\langle\beta\vert$. The Hamiltonian $\hat{H}$ in eq. (\ref{eq:ham}) consists of two parts. The first part is a local Hamiltonian $\hat{H}_i$ that contains the atom-light interactions. The second part describes the power-law interactions between two Rydberg atoms at sites $\textbf{r}_i = \left(x_i, y_i\right), x_i, y_i \in \left[1, L\right]$ and $\textbf{r}_j$ $(i\neq j)$ separated by a distance $\vert {\textbf r}_i - {\textbf r}_j\vert$. For a lattice spacing \textit{a} the nearest-neighbour coupling is $V_0 = C_{\alpha}/a^{\alpha}$ where $C_{\alpha} > 0$ determines the interaction strength. Dipole-dipole interactions are associated with $\alpha = 3$ and $\alpha = 6$ with van-der-Waals (vdW) interactions. In addition, the system undergoes decoherence due to strong spontaneous decay from state $\vert p \rangle$ at rate $\gamma_r$. We consider Markovian loss and decoherence described by the superoperator $\mathcal{L}\left[\rho\right]$. The N-body density matrix evolves as $\dot{\hat{\rho}} = -i\left[\hat{H}, \hat{\rho}\right] + \mathcal{L}\left[\hat{\rho}\right]$ where $\mathcal{L}\left[ \hat{\rho} \right] = \gamma_r \sum_i \left[\hat{\sigma}_{gp}^{(i)} \hat{\rho} \hat{\sigma}_{pg}^{(i) } - \frac{1}{2}\lbrace \hat{\sigma}_{pg}^{(i)}\hat{\sigma}_{gp}^{(i)}, \hat{\rho}\rbrace \right]$. 

For sufficiently strong decoherence, one can derive an effective rate equation as in subsection \ref{subsec:model} but for the three-level driving scheme by adiabatic elimination of the atomic coherences and neglecting multi-photon transition of two or more atoms \cite{Schonleber14,Ates2007,Ates2011,Heeg2012} \footnote{We have confirmed these simplifications via quantum simulations of smaller lattices.}. This simplifies the time evolution to the diagonal elements of $\rho$. The effective rate equation for the joint probabilities $\rho_{S_1,...,S_N}$ of Rydberg excitations being present ($S_i=1$) or not present ($S_i=0$) at the $i$th site and the corresponding many-body states are connected by the single-atom excitation [$\Gamma_{\uparrow}(\delta_i)$] and de-excitation [$\Gamma_{\downarrow}(\delta_i)$] rates as mentioned in (eq.~\ref{eq:rate_equ}). The specific form of those single-atom rates is given in \cite{Ates2007} and the single-atom steady state is a simple Lorentzian defined as
\begin{equation}\label{eq:p0}
\bar{\rho}_{1}(\delta)=\frac{\Gamma_{\uparrow}}{\Gamma_{\uparrow}+\Gamma_{\downarrow}}=\frac{p_0}{1+\delta^2/\omega^2}
\end{equation}
where $p_0$ denotes the resonant excitation probability with $p_0 = \Omega_1^2{\big /}\left( \Omega_1^2 +  \Omega_2^2\right)$ and $\omega$ denotes Lorentzian width with 

\begin{equation}
	\omega = \frac{\Omega_1^2+\Omega_2^2}{2\sqrt{\gamma_r^2+2\Omega_1^2}}.
\end{equation}
The interactions enter through an effective frequency detuning 
\begin{equation}
	\delta_i= \Delta-V_0 \sum_{j\neq i} \frac{S_j}{{\big |}{\bf r}_i-{\bf r}_j{\big |}^{\alpha}}
\end{equation}
which accounts for the level shift of the \textit{i}th atom due to its surrounding Rydberg excitations. All parameters are scaled by the Lorentzian width $\omega$, the many-body state is fully described by only four-parameter: the power law exponent $\alpha$, the resonant excitation probability $p_0$, laser detuning $\Delta / \omega$ and the interaction strength $V_0/\omega$. For the settings discussed in here [cf. Fig.\ref{fig1} (c)], the rates can be expressed as $\Gamma_\downarrow=(1-\bar{\rho}_{1})/T_1$, $\Gamma_\uparrow=\bar{\rho}_{1}/T_1$ where $T_1(\delta)$ denotes the onsite relaxation time.  

%%%%%%%%%%%%%%%%%%%%%%%%%%%%%%%%%%%%%%%%%
\begin{figure}[htbp]
   \centering
   \includegraphics[width = 100 mm]{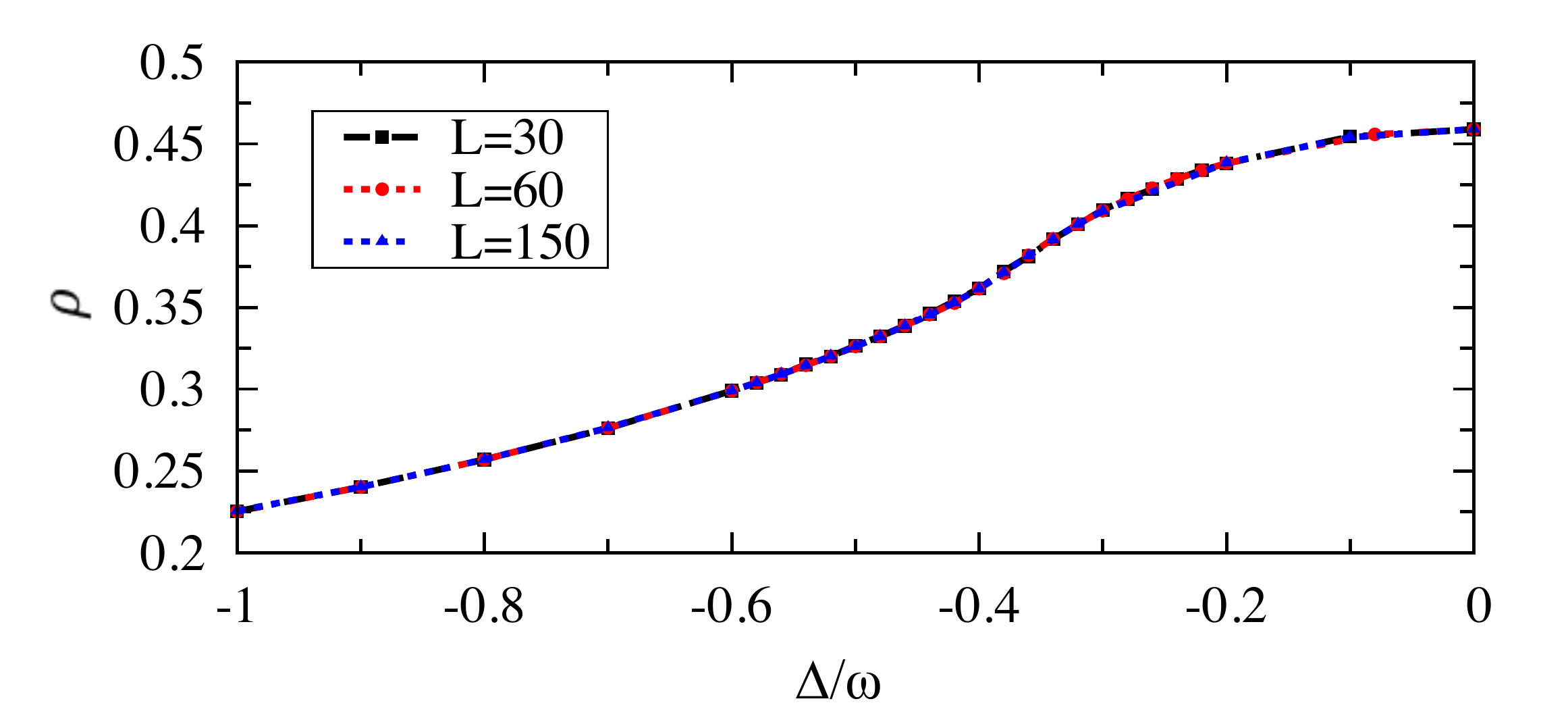} % requires the graphicx package
   \caption[q parameter on resonantly driven atoms]{The mean density as a function of laser detunings $\Delta$ for $p_0 \simeq 0.95$ and $V_0 = 5\omega$ calculated by steady-state Monte Carlo (ssMC). The simulations are performed with NN approximation and the symbols show results for finite system sizes L.}
   \label{fig:meanAF}
\end{figure}
%%%%%%%%%%%%%%%%%%%%%%%%%%%%%%%%%%%%%%%%%

The simulations have been performed by means of dynamic Monte Carlo (dMC) based on the rates $\Gamma_{\uparrow(\downarrow)}$ and steady-state Monte Carlo (ssMC), assuming $T_1(\delta)=T_1={\rm const.}$, and we have found good agreement in the relevant parameter regimes. We calculate the mean density as in section~\ref{sec:noise} for finite system sizes and nearest neighbours approximation as a function of the detuning $\Delta$, shown in Fig.~\ref{fig:meanAF}. We found that the mean density is independent of the system size. For detecting AF order, corresponding to a checkerboard configuration, we need to look at the populations at the lattice sites.   

\subsection{Long-range antiferromagnetic order}
\label{sec:2}
We define the order parameter $q$ that characterises the phase transition as

\begin{equation}
q=\frac{\vert N_e^A - N_e^B \vert}{N_e}
\end{equation}
where $N_e= \sum_{i}\langle \sigma_{ee}^{(i)}\rangle$. The excited state populations on the checkerboard sublattices A and B are denoted by $N_e^A$ and $N_e^B$, respectively. As illustrated in Fig.\ref{fig1}, $q$ measures the population imbalance on the two sublattices, with $q>0$ in the ordered phase and $q=0$ in the disordered phase that corresponds to a paramagnetic phase.

%%%%%%%%%%%%%%%%%%%%%%%%%%%%%%%%%%%%%%%%%
\begin{figure}[htbp]
   \centering
   \includegraphics[width = 100 mm]{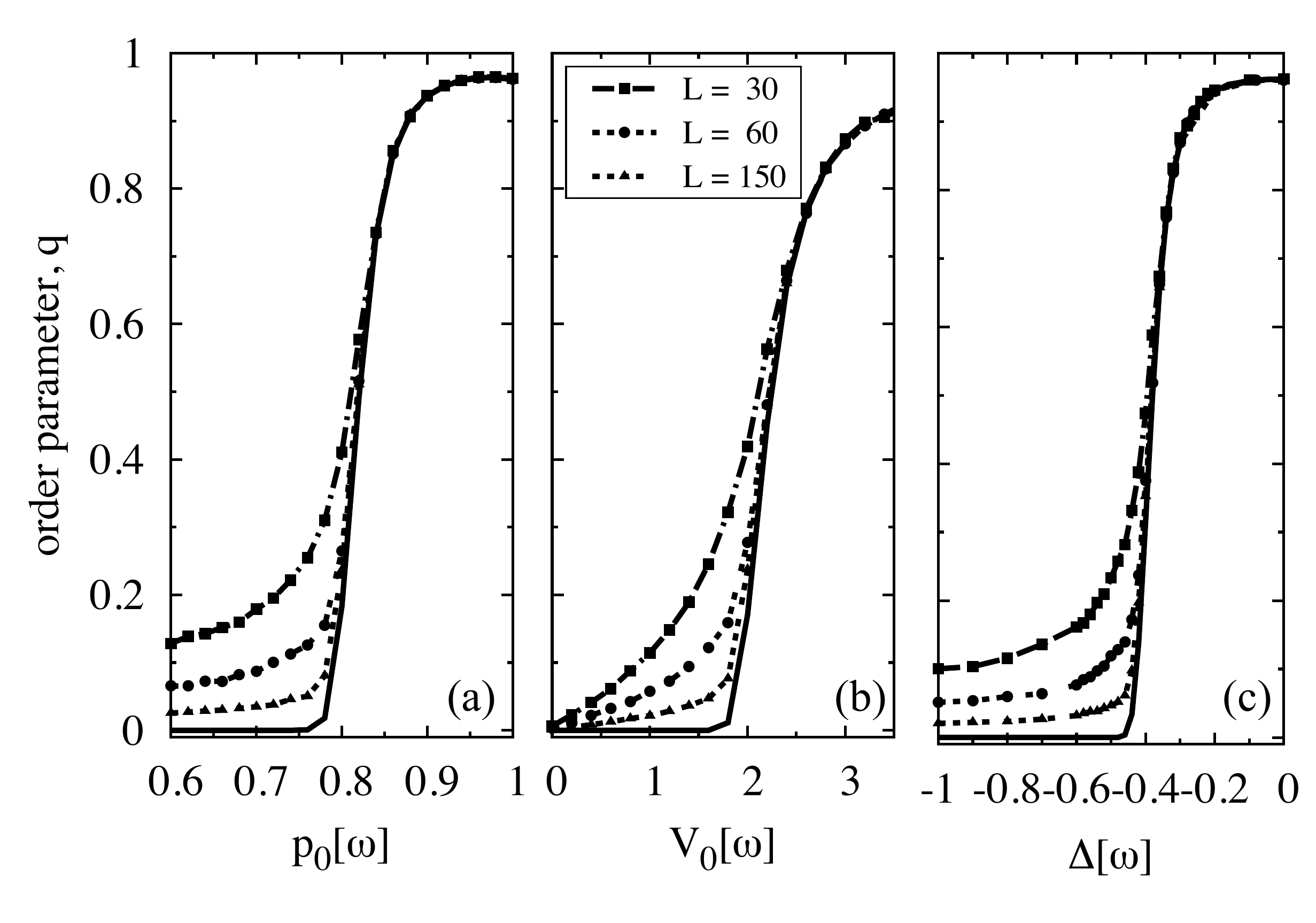} % requires the graphicx package
   \caption[q parameter on resonantly driven atoms]{Order parameter q as a function of the resonant excitation probability $p_0$ for $V_0 = 5\omega$ and $\Delta = 0$ (a). (b) shows q as a function of the interaction strength $V_0$ for $p_0 \simeq 0.95$ and $\Delta = 0$. (c) shows q as a function of the laser detuning $\Delta$ for $p_0 \simeq 0.95$ and $V_0 = 5\omega$. The simulations were performed with NN approximation and the symbols show results for finite system sizes given in the legend. The thick solid line shows the extrapolation to the thermodynamic limit, $L\rightarrow \infty$.}
   \label{fig:shortrange}
\end{figure}
%%%%%%%%%%%%%%%%%%%%%%%%%%%%%%%%%%%%%%%%%

\begin{figure}
   \centering
   \includegraphics[width = 100 mm]{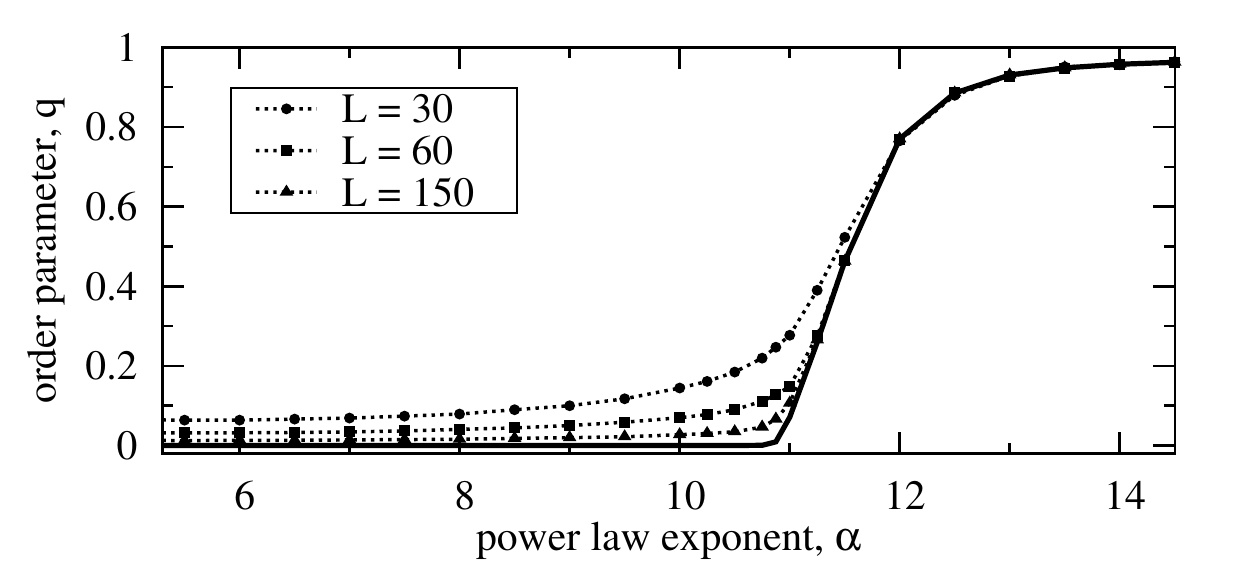} 
   \caption{Order parameter q as a function of the power-law exponent $\alpha$, for $p_0 = 0.95$ and $V_0 =  5 \omega$ and resonant driving $\Delta = 0$. The symbols correspond to different system sizes given in the legend. The thick solid line is the extrapolation to the thermodynamic limit, $L\rightarrow \infty$. The finite-size scaling shows a linear increase of the order parameter in the critical regime (exponent :  1 $\pm$ 0.05). (Figure adapted from~\cite{Hoening2014})}
   \label{fig2}
\end{figure}

For the one-dimensional lattice case, under the assumption of a NN-blockade, the above model is analytically solvable and shows no long-range order crystallisation. In higher dimensions, the steady states of N\'eel order occur for $p_0 \sim 0.7914$ in 2D and $p_0 \sim 0.749$ in 3D square lattices \cite{Pearce1988}. Thus, for simple two-level driving crystallisation is impossible in any dimension since $p_0\leq0.5$. We test our simulation in the NN-approximation and compare it to the analytical result, using ssMC simulations. As shown in Fig.\ref{fig:shortrange}, the steady state indeed exhibits N\'eel order provided that $p_0\sim 0.7914$  for $\Delta \approx 0$ and interactions. In the three-level scheme, by virtue of the dark state $D\sim\Omega_1 \vert e \rangle - \Omega_2 \vert g \rangle$, one can overcome the limit of two-level driving scheme $p_0\leq0.5$. However, when considering the full range power tail interactions for resonantly driven atoms with varying exponents $\alpha$ and $p_0=0.95$,  as shown in Fig.\ref{fig2}, we do not observe long-range order for realistic interaction potentials. In particular, the NN approximation fails qualitatively for the important case of vdW interactions ($\alpha=6$). Surprisingly, the weak tail of the interactions prevents crystallisation until a rather larger value $\alpha\approx11$. In fact, the simulations show that resonantly driven atoms, with vdW interactions remain in the disordered phase for any values of $p_0$ and $V_0$. 

%%%%%%%%%%%%%%%%%%%%%%%%%%%%%%%%%%%%%%%%%%%%%%%%%%%%%%%
%%%%%%%%%%%%%%%%%%%%%%%%%%%%%%%%%%%%%%%%%%%%%%%%%%%%%%%

\begin{figure}
   \centering
   \includegraphics[width = 100 mm]{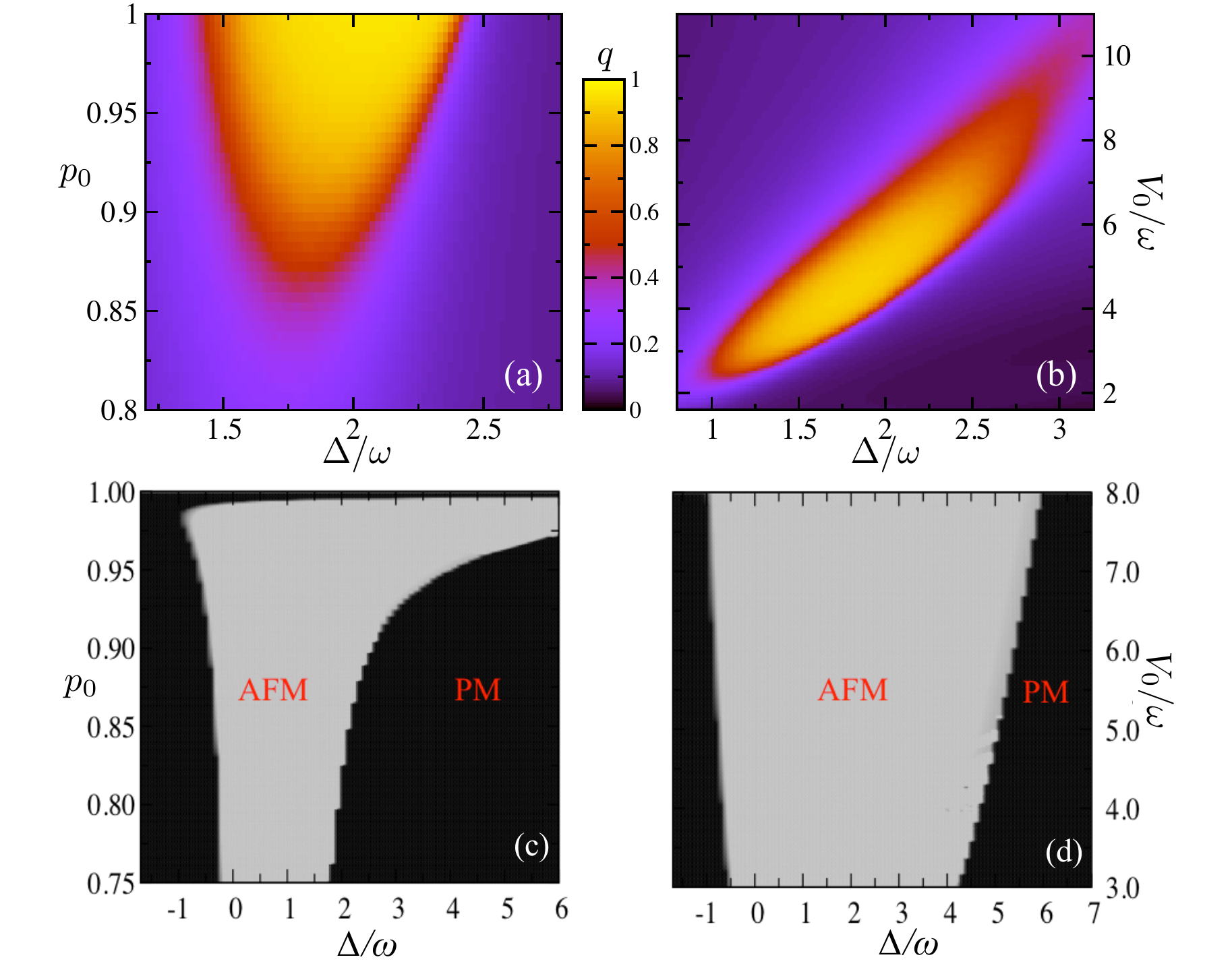} 
   \caption[Order parameters calculated with a Monte Carlo and meanfield methods.]{(Color online) Order parameter obtained for a 30 x 30 lattice as a function of $\Delta$ and $p_0$ for $V_0 = 5 \omega$ (a) and as a function of $\Delta$ and $V_0$ for $p_0=0.96$ (b) compared to mean-field predictions for the full vdW interactions (c) and (d). AFM and PM denote antiferromagnetic and paramagnetic order, respectively. [Figure (a) and (b) adapted from~\cite{Hoening2014}.]}
   \label{fig3}
\end{figure}

The fact that there is no phase transition on resonance for $\alpha = 6$ can be qualitatively understood as follows: a macroscopic population imbalance on the two sublattices each with a lattice constant $\sqrt{2}a$ characterises a N\'eel state. Assuming that an atom on the highly populated sublattice has an average $z$ nearest neighbours, the vdW interactions cause an energetic shift of $z V_0/8\omega$ and $z\approx3$ near the crystallisation transition. Therefore, the laser detuning $\Delta$ must compensate the corresponding energetic shift such that its excitation probability remains above threshold  $\bar{\rho}_1(\Delta/\omega-z V_0/8\omega)\geq p_{\rm c}$, with $\bar{\rho}_1$ given in eq.(\ref{eq:p0}). This is highlighted in Fig.\ref{fig3}, showing the order parameter $q$ for finite detunings $\Delta$, $p_0$ and $V_0$. As shown in Fig.\ref{fig3}(a), N\'eel-type ordering indeed emerges within a finite detuning range and for $p_0>p_{\rm c}\approx0.86$, only slightly larger than the threshold in the NN-blockade model \cite{Pearce1988}. Yet, N\'eel states are only found in a certain interval of interaction strengths $V_0$, since the vdW tail prevents long-range ordering beyond a critical value Fig.\ref{fig3}(b). The parameter region where this condition is fulfilled is marked in Fig.\ref{fig3}(a) and (b) and qualitatively reproduces our numerical results. In order to quantitatively assess the importance of fluctuations and the shape of the interaction potential, we have performed mean field simulations under the full vdW interactions. In opposite to the ssMC simulation, mean field predictions fail to show the threshold of $p_0$ in the NN-blockade model \cite{Pearce1988}, as shown in Fig.\ref{fig3}(c) and AF phase is found for any interaction strength $V_0$ [see Fig. \ref{fig3}(d)]. The phase transition between the AF and paramagnetic phase is a second order phase transition~\cite{Hoening2014}, in contrary to mean field and other predictions \cite{Weimer15,Overbeck16}, that suggest a first order transition. 

 %%%%%%%%%%%%%%%%%%%%%%%%%%%%%%%%%%%%%%%%%%%%%%%%%%%%%%%
%%%%%%%%%%%%%%%%%%%%%%%%%%%%%%%%%%%%%%%%%%%%%%%%%%%%%%%

\section{Experimental realisations}
\label{sec:exper}
We finally discuss an experimental realisation of two described systems above. For the Rydberg ensembles in the presence of laser phase noise described in section~\ref{sec:noise}, the dynamics of the Ising-like spin-1{\big /}2 system can be experimentally realised in a transverse field with up to thirty spins, for a variety of geometries in one and two dimensional lattices, and for a wide range of interaction strengths \cite{Labuhn15}. The interaction strength $V_0$ can be tuned easily by varying the lattice spacing $a > 3$ $\mu m$ and changing the principal quantum number \textit{n}, which scales approximately as $n^{11}$. As a specific example, laser excitation of Rb($58S_{1/2}$) Rydberg states with $\Omega/2\pi = 165 kHz$ and $\Gamma/2\pi = 0.7 MHz$ yields $\xi = 4$ \cite{Valado2015}. For a lattice constant of $a\approx 3.45 \mu $m these conditions corresponds to $R \approx 1.5$, i.e., well within the parameter region of strong interactions shown in Fig.~\ref{fig:dyn6atomsMC}(b)-(d). The interaction strength $R$ can be increased by tuning the laser excitation to higher principle quantum numbers, for example  Rb($67S_{1/2}$) yields $R \approx 2.0$ and Rb($75S_{1/2}$) yields $R \approx 2.5$. The speed up of the variance relaxation in the limit of weak interactions [see Fig.~\ref{fig:slopeQ}] can be observed when choosing larger lattice spacings $a$. For laser excitation of Rb($58S_{1/2}$), a lattice spacing in the range of $a\approx 8 - 14 \mu$m yields $R^6 \approx 0.01 - 0.35$, well within the parameter region shown in Fig.~\ref{fig:slopeQ} and \ref{fig:distribution_dyn}. The dependence of the steady-state variance on the number of atoms can be observed by adding up to 30 atoms.

In the presence of decoherence due to spontaneous emission,  the strong radiative decay of the intermediate state $\vert p \rangle$ with a rate $\gamma_r\sim$ MHz drives the relaxation towards the steady state eq.(\ref{eq:p0}), with a tuneable $p_0=\Omega_1^2/(\Omega_1^2+\Omega_2^2)$. Such three-level excitation schemes are utilised in numerous Rydberg atom experiments, either for exploring interaction effects in the strong excitation regime ($\Omega_1>\Omega_2$) \cite{Schempp2010,Schwarzkopf2013,Nipper2012,Viteau2012} or in quantum optics applications in the opposite limit \cite{Pritchard2010,Dudin2012,Peyronel2012,Maxwell2013}. As a specific example, laser excitation of Rb($35S_{1/2}$) Rydberg states via the intermediate Rb($5P_{1/2}$) state with $\Omega_1=0.5\gamma_r=4\Omega_2$ yields $p_0\approx0.9$. For a lattice constant of $a\approx2\mu$m these conditions correspond to $V_0\approx5\omega$, i.e. well within the parameter region of the ordered steady state. Rydberg excitation and trapping \cite{Anderson2011} as well as single-site resolved Rydberg atom imaging \cite{Schauss2012} have been experimentally demonstrated in 2D lattices with $a\approx0.5\mu$m. Larger lattice constants can also be realised in these settings \cite{Fukuhara2013} or via single-atom trapping in optical micro-trap arrays \cite{Beguin2013}, such that the creation and probing of the predicted dissipative phase transition appears to be well within experimental reach.

%%%%%%%%%%%%%%%%%%%%%%%%%%%%%%%%%%%%%%%%%\\
\section{Conclusions}
\label{sec:conclusion}
In conclusion, we have studied the differences and similarities between correlated and uncorrelated laser phase noise. Although the mean density in the steady state is identical, the excitation distributions differ considerably for both types of noise and in addition there is a dependence on the interaction strength. In the weak interaction limit, the relaxation of the mean density and variance for uncorrelated noise is independent of the interaction strength $R^6$, while for other limits it does depend on the interaction strength. In contrast to that, the relaxation in the presence of correlated noise shows a dependence for any interaction strength. We find a suppression of excitation fluctuations with increasing atom number in the presence of correlated noise, while for uncorrelated noise, the fluctuations are constant.  Current experiments \cite{Labuhn15,Beguin2013} are able to address this problem and could shed light on the nature of laser phase noise.  

In the presence of spontaneous decay, we have shown that Rydberg lattices can indeed undergo a dissipative phase transition to a long range ordered AF phase. The use of three-level schemes that go beyond the inversion limit of a simple two-level driving and a finite laser detuning are key requirements to counteract the effects of the power-law tail of the interaction potential. Dissipative phase transitions can also be realised in laser-cooled ion crystals in various spin models \cite{Porras2004} with $\alpha=0\ldots3$, in one and two dimensions \cite{Britton2012,Islam2013,Richerme2014}. In light of the demonstrated failure of mean field theory in \cite{Mendoza15,Weimer15,Maghrebi15}, it would further be of great interest to investigate other dissipative phase transitions predicted by mean field treatment \cite{Lee2013,Qian15,Chan15} and gain insight into their validity for open systems as well as the critical dimension for long-range order in such related spin lattices. 

\section*{Acknowledgements}
We thank M. H\"oning, M. Fleischhauer and T. Pohl, with whom parts of the work discussed in section~\ref{sec:1} have been performed. We acknowledge useful discussions with I. Lesanovsky and A. Eisfeld. We thank T. Pohl for valuable feedback and guidance. Financial support by the EU through the Marie Curie ITN "COHERENCE" is acknowledged.

\end{document}